\documentclass{optica-article}
\journal{opticajournal} 
\articletype{Research Article}
\usepackage{lineno}
\usepackage{dcolumn}
\usepackage{braket}
\usepackage{mathtools}
\usepackage{graphicx}
\usepackage{mathrsfs}
\usepackage{mdwlist}
\usepackage{subfigure}
\usepackage{booktabs}
\usepackage{amsmath}
\usepackage{appendix}
\usepackage{extarrows}
\usepackage{dsfont}
\usepackage{amstext}
\usepackage{amsbsy}
\usepackage{bbm}
\usepackage{bm}
\usepackage{graphicx}
\usepackage{textcomp}
\usepackage{multirow}
\usepackage{threeparttable}

\usepackage{color}
\usepackage{diagbox}
\definecolor{Dgreen}{RGB}{0, 100, 0}
\usepackage{url}

\begin{document}

	\title{{Optimal robust control of cat-state qubits against parameter imperfections}}
	\author{Shao-Wei Xu,\authormark{1,2} Zhe-Yuan Zhang,\authormark{2} Jiang-Ting Ye,\authormark{2} Zhong-Zheng Zhang,\authormark{2} Yue-Ying Guo,\authormark{2} Ke-Xiong Yan,\authormark{1,2} Ye-Hong Chen,\authormark{1,2,3,5} and \author{Yan Xia\authormark{1,2,4}}}
	
	\address{\authormark{1}Fujian Key Laboratory of Quantum Information and Quantum Optics (Fuzhou University), Fuzhou 350116, China\\
		\authormark{2}Department of Physics, Fuzhou University, Fuzhou 350116, China\\
		\authormark{3}Theoretical Quantum Physics Laboratory, Cluster for Pioneering Research, RIKEN, Wako-shi, Saitama 351-0198, Japan\\
	\authormark{4}Institute of Quantum Science and Technology, Yanbian University, Yanji, Jilin 133002, China}
		
	\email{\authormark{5}yehong.chen@fzu.edu.cn}

    \begin{abstract}
        Cat-state qubits formed by photonic coherent states are a promising candidate for realizing fault-tolerant quantum computing. Such logic qubits have a biased noise channel that the bit-flip error dominates over all the other errors. 
        In this manuscript, we propose an optimally robust protocol using the control method of shortcuts to adiabaticity to 
        realize a high-fidelity {state transfer} in a cat-state qubit. 
        We construct a shortcut based on the Lewis-Riesenfeld invariant and examine the stability versus
        different types of perturbations for the fast and robust {bit flipping}. 
        Numerical simulations demonstrate that the {bit flipping} can be {robust against}
        systematic errors in our protocol. Even when the parameter imperfection rate for bit-flip control is $20\%$, 
        the final population of the target state can still reach $\geq 99\%$. 
        The optimally robust control provides a
        feasible method for fault-tolerant and scalable quantum computation.
    \end{abstract}
    \date{\today}

    \section{Introduction}
    Quantum computers promise to drastically outperform classical computers on certain problems, such as factoring, (approximate) optimization, boson sampling, or
    unstructured database searching \cite{Scully1997Book,Agarwal2012Book,HidaryBook,LiptonBook,Kockum2019}. Building a large-scale quantum computer requires qubits that can  be protected from errors, i.e., utilizing quantum 
    error correction. During the past decades, many strategies using physical and logical qubits for quantum error correction have been developed. Noting that 
    quantum error correction with physical qubits usually requires huge physical resource overhead, this makes it difficult to scale up the number of 
    qubits for a large-scale quantum computer \cite{Shor1995,Steane1996,Lidar2013,Kjaergaard2020,Kitaev2003,Cai2021Fr,Ma2021Sb}. This is why in recent years, much attention has been paid to logic qubits formed by bosonic codes \cite{Ralph2003,Mirrahimi2014,Mirrahimi2016,Chamberland2020,Cai2021Fr,Ma2021Sb}, which 
    allow quantum error correction extending only the number of excitation instead of the number of qubits.
    
    A promising alternative with the potential to realize quantum error correction beyond the break-even point involves encoding logical qubits in continuous 
    variables \cite{Aliferis2008,Mirrahimi2014,Mirrahimi2016,Puri2017npjQI,Puri2019Prx,Puri2020Sa,Albert2016,Albert2019,Cai2021Prl,Zheng2023Prl,Li2024Pra}, such as coherent states. This gives rise to the cat-state codes, which are formed by even and odd coherent states of a single optical
    mode \cite{Mirrahimi2014,Mirrahimi2016,Puri2017npjQI,Puri2019Prx,Puri2020Sa,Albert2016,Albert2019,Chen2021,Chen2022Prappl,Chen2024Prl,Zhao2025OE}. The cat-state qubits preserve the noise bias that experience only bit-flip noise, reducing the number of building blocks of layers for error 
    correction \cite{Mirrahimi2016,Puri2017npjQI,Puri2019Prx,Puri2020Sa,Chen2024Prl}. Moreover, the first experiment \cite{Grimm2020} realizing cat-state qubits showed a strong suppression of frequency fluctuations due to $1/f$ noise \cite{Grimm2020,Darmawan2021PrxQ,Xu2022book,Kang2022Prr}. 
    All these make the cat-state qubits promising for hardware efficient universal quantum computing.
    
    In an implementation of quantum computation, high-fidelity single- and two-qubit quantum gates are essential elements of quantum computation because quantum 
    algorithms are usually designed as a sequence of such simple quantum gates \cite{Kockum2019,Xu2022book}. Though several experiments have realized the control of cat-state qubits \cite{Grimm2020,WangPrx2019}, the robust control of a single cat-state qubit is still a problem to be 
    solved. In this manuscript, we propose an optimally robust shortcuts to adiabatic protocol for controlling a cat-state qubit. Shortcuts to adiabaticity \cite{Chen2010Prl,Chen2010Prl,Campbell2015Prl,Torrontegui2013Aamop,Zhou2016Np,Du2016Nc,Hatomura2018Njp,Ibanez2012Prl,Guery2019Rmp,Funo2020Prl,Abah2020Prl,Chen2021,Takahashi2024Prx} 
    are a series of protocols mimicking adiabatic dynamics beyond the adiabatic limit and have been widely applied for quantum state engineering. One of 
    the more prominent of these protocols is the method of ``invariant-based reverse engineering'' \cite{Lewis1969Jmp,Chen2011Pra}, which can construct shortcuts only by redesigning 
    system parameters without destroying the initial form of the system Hamiltonian. This provides an alternative control method for the cat-state qubits with 
    large amplitudes because such qubits can be manipulated along only one direction on the Bloch sphere \cite{Puri2019Prx}. Moreover, the invariant-based reverse engineering 
    is compatible with various quantum optimized control techniques \cite{Ruschhaupt2012Njp}. One can thus optimize the parameters to realize a high-fidelity {bit flipping} 
    of a cat-state qubit.
    
    This manuscript is organized as follows. 
	{In Sec. \ref{sec2}, we present a model to stabilize cat-qubits by using Kerr-nonlinear resonator and derive the effective Hamiltonian for the 
	protocol.} 
	The protocol of constructing shortcuts to adiabatic passage is given in Sec.~\ref{sec3}.
    In Sec.~\ref{sec5}, we analyze the systemic error sensitivity of the cat-state qubit. Then, in Sec.~\ref{sec6} the optimal protocol to minimize the systemic error sensitivity is presented. Moreover, we discuss the influence of single-photon loss and pure dephasing on the protocol in Sec.~\ref{sec8}. Finally,
    the conclusions are given in Sec.~\ref{sec9}.
    
    \begin{figure}
    	\centering
    	\scalebox{0.45}{\includegraphics{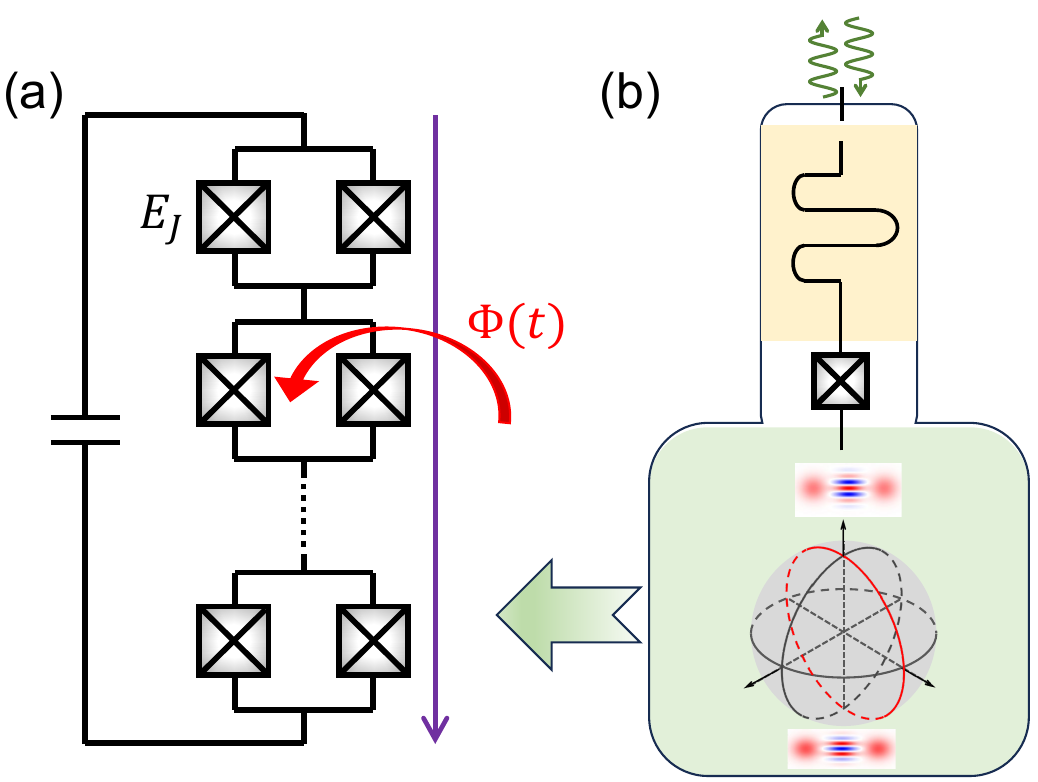}}
    	\caption{Schematic of the realization of a controllable single-mode cat-state qubit. (a) An array of Josephson junctions for modeling a Kerr-nonlinear resonator, where the Josephson
    		energy $E_J$ is tunable by controlling the external magnetic flux $\Phi(t)$. (b) We couple the cat-state qubit to a low-Q readout mode through a high-impedance
    		Josephson circuit.}
    	\label{fig0}
    \end{figure}

    \section{Model and effective Hamiltonian}\label{sec2}
    {We consider a model with a Kerr-nonlinear resonator \cite{Puri2019Prx,Mirrahimi2014NJP, Puri2017npjQI, Grimm2020, Gautier2022PRXQ, Ruiz2023PRA}.} 
	The Kerr-nonlinear resonator with frequency $\omega_{c}$ is driven by a single-mode, two-photon excitation \cite{Chen2021,Yu2025OE}, where the driving frequency for the two-photon 
	excitation is twice the resonator frequency. In the rotating-wave approximation, the system Hamiltonian is given by (hereafter $\hbar=1$)
    \begin{align}\label{eq1}
        {H}_{\mathrm{Kerr}} = -K {a}^{\dagger 2} {a}^{2} + P({a}^{\dagger2} + {a}^{2}). 
    \end{align}
    In the above expression, ${a}$ and ${a}^{\dagger}$ are the annihilation and creation operators for photons, $K$ is the strength of the 
    Kerr-nonlinearity, and $P$ is the strength of the two-photon drive. 
    {Such a model can be realized using an array of Josephson junctions, in which the Josephson
    energy $E_J$ is tunable by controlling the external magnetic flux $\Phi(t)$ as shown in Fig.~\ref{fig0}(a) \cite{Chen2022Prappl,Puri2019Prx,WangPrx2019}. The standard quantization procedure for the circuits are given in Appendix \ref{secA1}.}
    
\begin{figure}
        \centering
        \scalebox{0.35}{\includegraphics{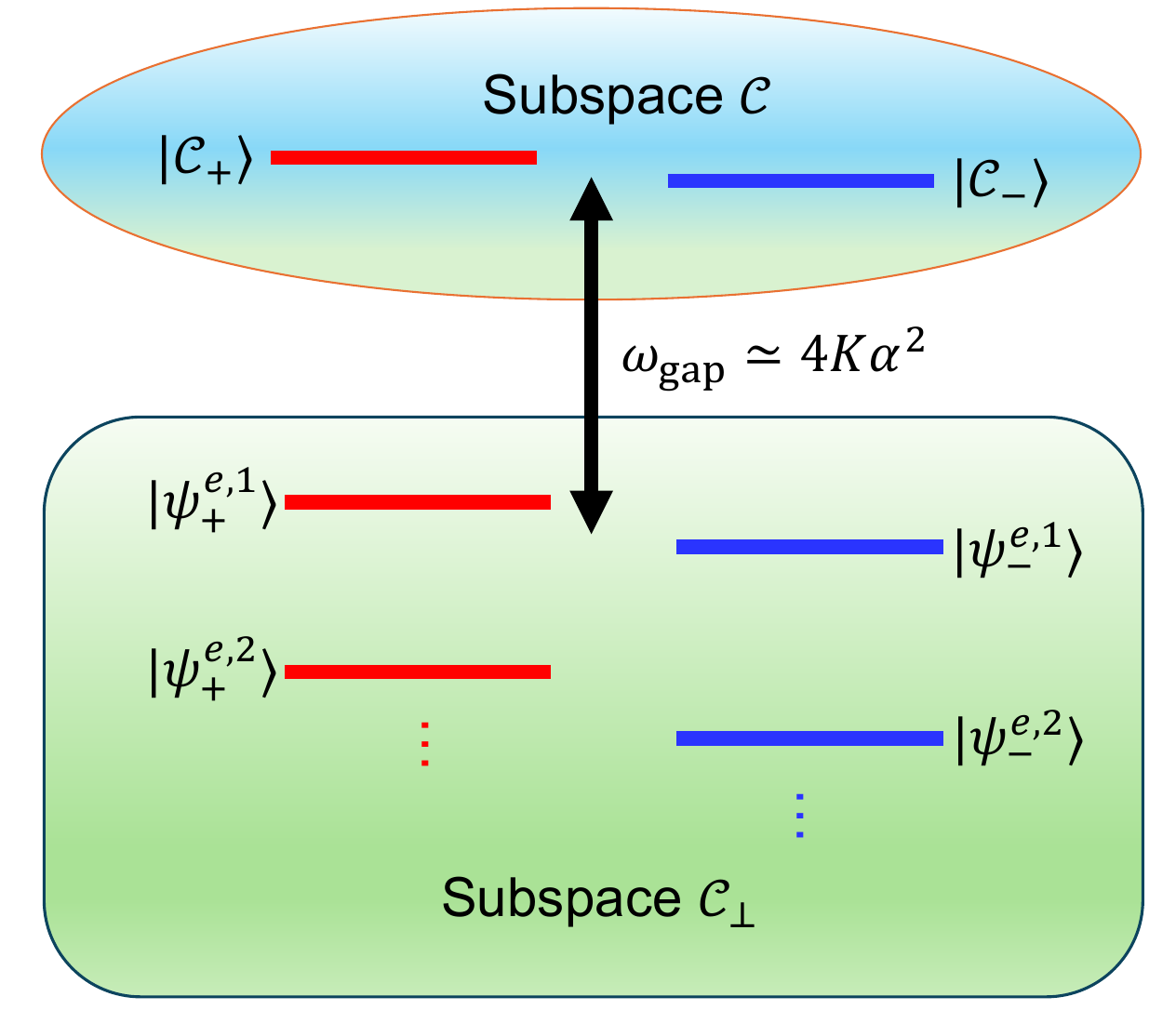}}
        \caption{In the rotating frame determined by Eq.~(\ref{eq2}), the characteristic spectrum of the Kerr-nonlinear resonator ${H}_{\mathrm{Kerr}}$.}
        \label{fig1}
    \end{figure}

    We can observe that Eq.~(\ref{eq1}) is written in the rotating frame. In this frame, the 
    simplified Hamiltonian is described as having quasi-energy eigenstates with negative energies. Specifically, by applying the displacement transformation 
    $D(\pm \alpha) = \operatorname{exp} [\pm \alpha ({a}^{\dagger} - {a}) ]$ to ${H}_{\mathrm{Kerr}}$, the Hamiltonian in Eq.~(\ref{eq1}) becomes 
    \begin{align}\label{eq2}
        {H}^{\prime} &= D(\pm \alpha) {H}_{\mathrm{Kerr}} D^{\dagger}(\pm \alpha) \cr\cr
                         &=-4K \alpha^{2} {a}^{\dagger} {a} - K {a}^{\dagger 2} {a}^{2} \mp 2K \alpha ({a}^{\dagger 2} {a} + \mathrm{H.c.}),
    \end{align}
    where $\alpha = \sqrt{\frac{P}{K}}$. The vacuum $|0\rangle$ is exactly an eigenstate of ${H}^{\prime}$. Therefore, the coherent states 
    $D(\pm\alpha)|0\rangle=|\pm \alpha \rangle$ or, equivalently, their superposition states 
    \begin{align}\label{eq3}
    	 | \mathcal{C}_{\pm} \rangle = {N}_{\pm} (| \alpha \rangle \pm | -\alpha \rangle),
    \end{align}
    are the degenerate eigenstates of ${H}_{\mathrm{Kerr}}$, where ${N}_{\pm} = 1 / \sqrt{2(1 \pm e^{-2|\alpha|^2})}$ are normalized coefficients. 
    In the limit of large $\alpha$, one can obtain $\alpha^2 \gg \alpha^1, \alpha^0$. Thus, Eq.~(\ref{eq2}) is approximated by 
    ${H}^{\prime} \simeq -4K \alpha^{2} {a}^{\dagger} {a}$, which is the Hamiltonian of a (inverted) harmonic oscillator. Therefore, in the original 
    frame, the eigenstates of ${H}_{\mathrm{Kerr}}$ are eigenstates of the parity operator (see Fig.~\ref{fig1}). The first-excited states can be approximately 
    expressed as two orthogonal states $|\psi_{\pm}^{e,1} \rangle = N_{\pm}^{e,1} [D(\alpha) \pm D(-\alpha)] |n=1\rangle$, where $N_{\pm}^{e,1}$ are normalized 
    coefficients and $|n\rangle$ are Fock states. The energy gap between the cat states subspace $\mathcal{C}$ and $|\psi_{\pm}^{e,1} \rangle$ can be approximated as 
    $\omega_{\mathrm{gap}} \simeq 4K \alpha^2$.
        
    As shown in Fig.~\ref{fig1}, the cat states subspace $\mathcal{C}$ is separated from the rest of the Hilbert space $\mathcal{C}_\perp$ by an energy gap of 
    approximately $\omega_{\mathrm{gap}} \simeq 4K \alpha^2$. The action of ${a}$ can only flip the two cat states 
    $| \mathcal{C}_{\pm} \rangle$, i.e.,
    {\begin{align}\label{eq4}
        {a} | \mathcal{C}_{-} \rangle=&\alpha\sqrt{\tanh|\alpha|^{2}}|\mathcal{C}_{+} \rangle,\cr
          {a} | \mathcal{C}_{+} \rangle=&\alpha \sqrt{\coth|\alpha|^{2}}|\mathcal{C}_{-} \rangle. 
    \end{align}Thus, in the limit of large $\alpha$, we have $\tanh|\alpha|^2\simeq\coth{|\alpha|^{2}}\simeq 1$, resulting in
	$a|\mathcal{C}_{\pm}\rangle\simeq\alpha|\mathcal{C}_{\mp}\rangle$}.
 	{
	 
	Based on the invariant-based reverse engineering \cite{Chen2011Pra,Ruschhaupt2012Njp}, we introduce a quantum control strategy which 
	ensures a high-fidelity {state transfer} while minimizing sensitivity to systematic errors. Therefore, in the interaction 
	picture, we add a control Hamiltonian \cite{Cohen2017Prl, Chen2010Prl, Kang2022Prr}}
    \begin{align}\label{eq5}
    	H_{c}(t)=&-\frac{E_{J}(t)}{2}\{D[i\varphi_{a}\exp(i \omega_{c}t)]+{\rm H.c.}\}\cr
    	      &+\epsilon(t) ({a}^{\dagger}+{a}).
    \end{align}
    A possible implementation of this control Hamiltonian is the superconducting circuits \cite{Cohen2017Prl}
    by {capacitively coupling} the Kerr-nonlinear mode to a Josephson junction and assuming that other modes (including the junction mode) are never excited.
    Accordingly, the time-dependent parameter $E_{J}(t)$ is the effective Josephson energy and $\varphi_{a}$ is the phase, 
	{see Appendix~\ref{EJ} for more details.} A single photon driving with time-dependent amplitude $\epsilon(t)$ is also applied to the system.
    For $E_{J}(t),\epsilon(t)\ll\omega_{c}$, the { control} Hamiltonian under the rotating-wave approximation becomes
    \begin{align}\label{eq6}
        {H}'_{c}(t) = E_{J}e^{-\varphi_{a}^{2}/2}\sum_{m=0}^{\infty}L_{m}(\varphi_{a}^{2})|m\rangle\langle m| + \epsilon ({a}^{\dagger}+{a}),
    \end{align}
    where $L_{m}(*)$ is the Laguerre polynomial of order $m$.  Hereafter, for simplicity, we omit the explicit time
    dependence of parameters, e.g., $E_{j}(t)\rightarrow E_{J}$ and $\epsilon(t)\rightarrow \epsilon$.

\begin{figure}
	\centering
	\scalebox{0.7}{\includegraphics{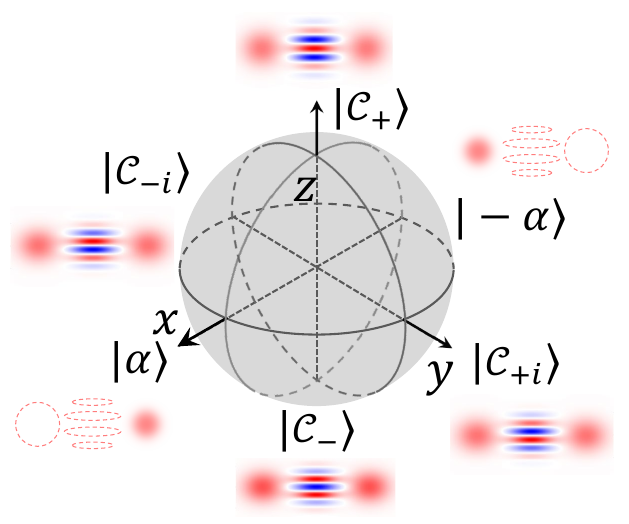}}
	\caption{Bloch sphere of the cat-state qubit described by Eq.~(\ref{eq10}) when $\alpha=2$. For simplicity, we assume $\alpha=2$ through out the manuscript.}
	\label{fig1b}
    \end{figure}

    The total Hamiltonian now becomes
    ${{H}_{\rm tot}(t)} = {H}_{\mathrm{Kerr}} + {H}'_{c}(t)$. We can use the cat states $| \mathcal{C}_{\pm} \rangle$ to define the Pauli matrices, 
    \begin{align}\label{eq7}
        &\sigma_{x} = \sigma_{+} + \sigma_{-},\cr 
        &\sigma_{y} = i(\sigma_{-} - \sigma_{+}),\cr 
        &\sigma_{z} = \sigma_{+} \sigma_{-} - \sigma_{-} \sigma_{+}, 
    \end{align}
    where $\sigma_{+} = | \mathcal{C}_{+} \rangle \langle \mathcal{C}_{-} |$ is the raising operator and 
    $\sigma_{-} = | \mathcal{C}_{-} \rangle \langle \mathcal{C}_{+} |$ is the lowering operator. When
    \begin{align}\label{eq8}
        E_{J},\ \epsilon \ll \omega_{\mathrm{gap}}, 
    \end{align}
    the evolution of the system can be restricted to the cat-state subspace $\mathcal{C}$, i.e., constructing a
   cat-state qubit as shown in Fig.~\ref{fig1b}. 

   {Projecting the system onto the cat-state subspace, the effective part of the total Hamiltonian $H_{\mathrm{eff}}$ can be represented as}
    \begin{align}\label{eq9}
            {H}_{\mathrm{eff}}
            =\frac{E_{J}\exp\left[-(\varphi_{a}-2\alpha)^2/2\right]}{-2\sqrt{\pi\alpha\varphi_{a}}} \sigma_z + \epsilon (\alpha^* + \alpha) \sigma_x.
    \end{align}
    We can choose $\varphi_{a}=2\alpha$ and rewrite the effective Hamiltonian in the matrix form as
    \begin{align}\label{eq10}
        {H}_{\mathrm{eff}} = \frac{1}{2}
        \begin{pmatrix}
            \Delta& \Omega_R\\
            \Omega_R & -\Delta
        \end{pmatrix},
    \end{align}
    where the time-dependent parameters are $\Delta =-E_{J}/(\alpha\sqrt{2\pi})$ and $\Omega_R = 2 (\alpha^* + \alpha) \epsilon$.   

    \section{Non-adiabatic evolution based on the Lewis-Riesenfeld invariants}\label{sec3}
    Following the method of invariant-based reverse engineering \cite{Lewis1969Jmp,Chen2011Pra}, we introduce a dynamical invariant $I(t)$, which satisfies
    \begin{align}\label{eq11}
    	i\frac{\partial}{\partial t}I(t)-[H_{\rm eff}(t),I(t)]=0.
    \end{align}
    Then, the solution of the time-dependent Schr\"{o}dinger equation
    \begin{align}\label{eq12}
    	i\frac{\partial}{\partial t}|\psi(t)\rangle=H_{\rm eff}(t)|\psi(t)\rangle,
    \end{align}
    can be expressed by a superposition of the eigenstates $|\phi_{n}(t)\rangle$ of $I(t)$ as
    \begin{align}\label{eq13}
	    |\psi(t)\rangle=\sum_{n}c_{n}\psi_{n}(t).   
    \end{align}
    Here, $\psi_{n}(t)=e^{iR_{n}(t)}|\phi_{n}(t)\rangle$ and $c_{n}$ are time-independent amplitudes determined by the initial state, and $R_{n}(t)$ are the 
    Lewis-Riesenfeld phases defined as
    \begin{align}\label{eq14}
	    R_{n}(t)=\int_{0}^{t}\langle \phi_{n}(t')|i\frac{\partial}{\partial t'}-H_{\rm eff}(t')|\phi_{n}(t')\rangle dt'.
    \end{align}

    Following {Refs.}~\cite{Chen2011Pra,Chen2010Prl,Ruschhaupt2012Njp},  for the Hamiltonian in Eq.~(\ref{eq10}), the invariant $I(t)$ can be given by 
    \begin{align}\label{eq15}
        I(t)=\frac{1}{2}
        \begin{pmatrix}
  	        \cos{\gamma} & \sin\gamma e^{i\beta}\\
  	        \sin\gamma e^{-i\beta} & -\cos{\gamma}
        \end{pmatrix}, 
    \end{align}
    where $\gamma$ and $\beta$ are two time-dependent dimensionless parameters to be determined later. The eigenstates of the Lewis-Riesenfeld invariant $I(t)$ 
    can be thus derived as \cite{Chen2011Pra}
    \begin{align}\label{eq16}
        | \phi_{+}(t) \rangle &= \cos{(\frac{\gamma}{2})} e^{i \beta} | \mathcal{C}_{-} \rangle + \sin{(\frac{\gamma} {2})} | \mathcal{C}_{+} \rangle, \cr\cr
        | \phi_{-}(t) \rangle &= \sin{(\frac{\gamma}{2})} | \mathcal{C}_{-} \rangle - \cos{(\frac{\gamma} {2})} e^{-i \beta} | \mathcal{C}_{+} \rangle. 
    \end{align}
    According to Eq.~(\ref{eq11}), we obtain the expressions of the time-dependent parameters $\Omega_R$ and $\Delta$ as 
    \begin{align}\label{eq17}
        \Omega_R &= \dot{\gamma} / \sin{\beta},  \cr
        \Delta &= \Omega_R \cot{\gamma} \cos{\beta} - \dot{\beta}. 
    \end{align}
    To achieve the flipping of the cat states $| \mathcal{C}_{\pm} \rangle$, one needs to set the boundary conditions $\Omega_R(0) = \Omega_R(t_f) = 0$, and 
    \begin{align}\label{eq18}
        &\gamma(0) = \pi,\ \ \ \ \gamma(t_f) = 0, \cr
        &\dot{\gamma}(0) = 0,\ \ \ \ \dot{\gamma}(t_f) = 0.
    \end{align}
    We can arbitrarily choose the values of $\beta(0)$ and $\beta(t_f)$, according to Eq.~(\ref{eq17}), when $\beta$ approaches $(n+1/2)\pi$, the resulting of 
    $|\Omega_R|$ is minimized, imposing 
    \begin{equation}\label{eq19}
        \begin{array}{c}
            \beta(0) = -\pi / 2,\ \beta(t_f/2 ) = -\pi / 2,\ \beta(t_f) = -\pi / 2,\\
            \dot{\beta}(0) = \pi / (2t_f),\ \dot{\beta}(t_f) = \pi / (2t_f).
        \end{array} 
    \end{equation}

\begin{figure}
	\centering
	\scalebox{0.42}{\includegraphics{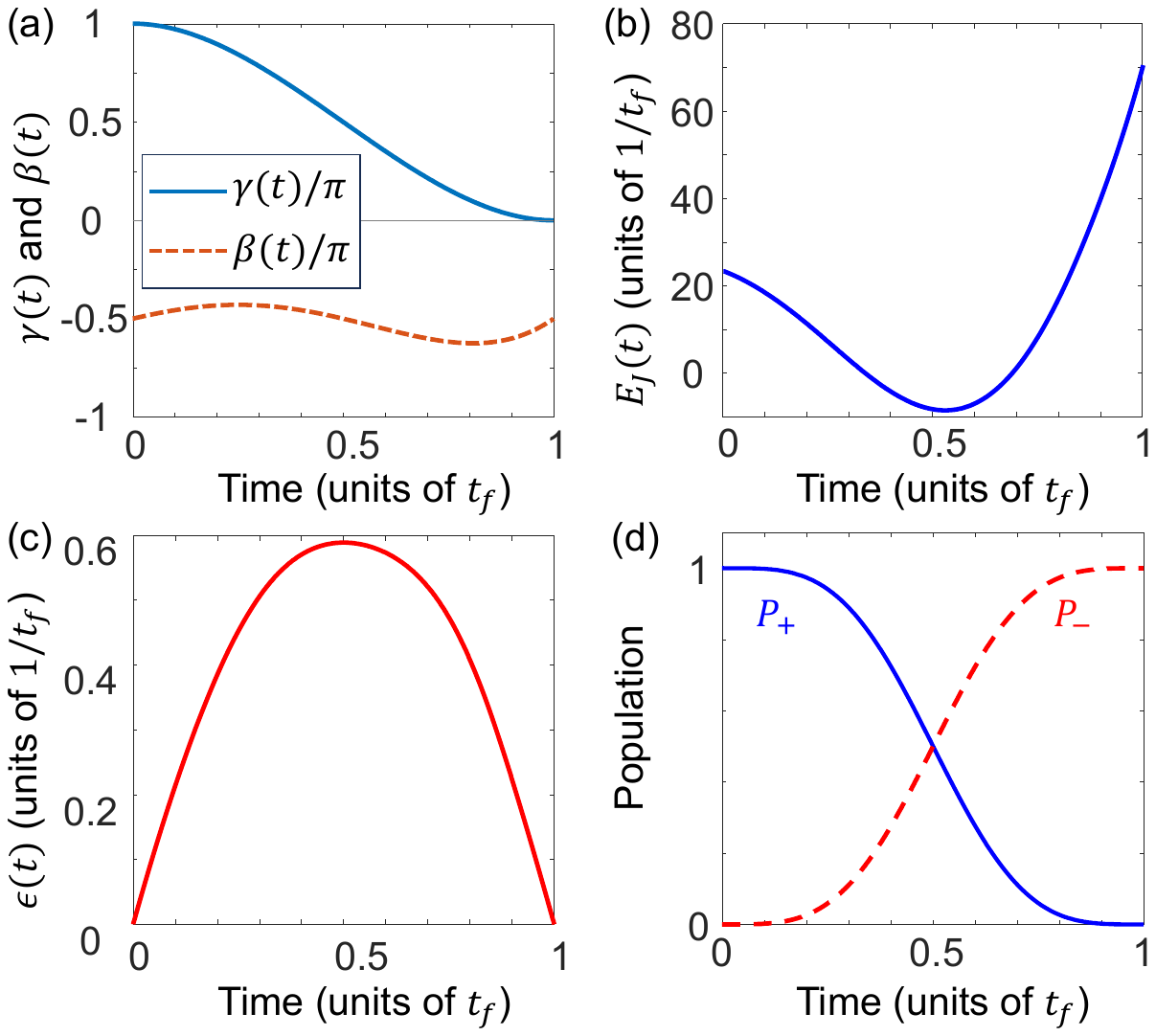}}
	\caption{(a) Polynomials $\gamma(t) = \sum_{i=0}^{3} a_i t^i$ (solid-blue curve) and $\beta(t) = \sum_{i=0}^{4} b_i t^i$ (dashed-orange curve). 
		(b) Corresponding function of $E_J$ calculated by $\Delta$ in Eq.~(\ref{eq17}).
		(c) Corresponding function of $\epsilon$ calculated by $\Omega_{R}$ in Eq.~(\ref{eq17}).
		(d) Non-adiabatic {bit flipping} in the cat-state qubit. We choose $t_{f}=5/K$ to satisfy the
		condition $E_{J},\epsilon\ll \omega_{\rm gap}$.
		{The evolution takes place in the complete Hilbert space.}
		}
	\label{fig2}
\end{figure}   
  
    To satisfy the boundary conditions given in Eqs.~(\ref{eq18}) and (\ref{eq19}), we can assume
    \begin{align}\label{eq20}
      \gamma(t) = \sum_{i=0}^{3} a_i t^i, \ \ \ \ \  {\rm and}  \ \ \ \  \ \beta(t) = \sum_{i=0}^{4} b_i t^i,
    \end{align}
    and thus determine their values as shown in Fig.~\ref{fig2}(a). Accordingly, we can obtain $E_J$ and $\epsilon$ [see Fig.~\ref{fig2}(b) and 
    Fig.~\ref{fig2}(c)]. Such parameters allow a {state transfer} from the even cat state
    $|\mathcal{C}_{+}\rangle$ to the odd cat state $|\mathcal{C}_{-}\rangle$ through a nonadiabatic passage. This is determined 
    by solving the Schr\"{o}dinger equation $i|\dot{\psi}(t)\rangle= {H_{\rm tot}} |\psi(t)\rangle$ of the total Hamiltonian 
    \begin{align}\label{eq21}
    	{H_{\rm tot}}=H_{\rm Kerr}+H'_{c}(t).
    \end{align}
    In Fig.~\ref{fig2}(d), we display the dynamical evolution of the system when the initial state is $|\mathcal{C}_{+}\rangle$.
    An almost perfect {bit flipping} ($P_{-}\simeq 99.9\%$ at $t=t_{f}$) is obtained as shown in the figure, where
    the populations of the states $|\mathcal{C}_{+}\rangle$ and $|\mathcal{C}_{-}\rangle$ are defined
    as 
    \begin{align}\label{eq22}
      P_{\pm}(t)=|\langle C_{\pm}|\psi(t)\rangle|^{2}.
    \end{align}

    {What's more, we can achieve the flipping of the coherent states $|\pm \alpha \rangle$ by redesigning the boundary conditions, see Appendix 
	\ref{Re} for more details.}

    \section{Systematic error sensitivity}\label{sec5}
    Now, we consider the influence of systematic errors on the system dynamics. The ideal undisturbed Hamiltonian is ${H}_{\mathrm{eff}}$. For systematic errors, the Hamiltonian in 
    actual experiments becomes ${H}_{01} = {H}_0 + \mu{H}_1$, which also satisfies the Schr\"{o}dinger equation
    \begin{align}\label{eq23}
        i \frac{d}{dt} | \psi(t) \rangle = ( {H}_0 + \mu {H}_1 ) | \psi(t) \rangle, 
    \end{align}
    where ${H}_0 = {H}_{\mathrm{eff}}$, and ${H}_1$ is the disturbed Hamiltonian. 
   For simplicity, we assume that errors affect the pulse amplitude but not the
   detuning. The disturbed Hamiltonian $H_{1}$ under this assumption is
   \begin{align}\label{eq24}
   	  H_{1}=\epsilon(\alpha^{*}+\alpha)\sigma_{x}.
   \end{align}
In the original control Hamiltonian $H_{c}$, this disturbed Hamiltonian corresponds to parameter deviations in
the single-photon drive $\epsilon(a+a^{\dag})$.
   Then, we define the systematic error sensitivity as
    \begin{align}\label{eq25}
        q_s = -\frac{1}{2} \frac{\partial^2 P_-^2}{\partial \mu^2} \Bigg|_{\mu=0} = -\frac{\partial P_-}{\partial( \mu^{2} )} \Bigg|_{\mu=0}, 
    \end{align}
    where $P_-$ is the population of the state $| \mathcal{C}_{-} \rangle$ at the final time $t_f$.
    
\begin{figure}
	\centering
	\scalebox{0.33}{\includegraphics{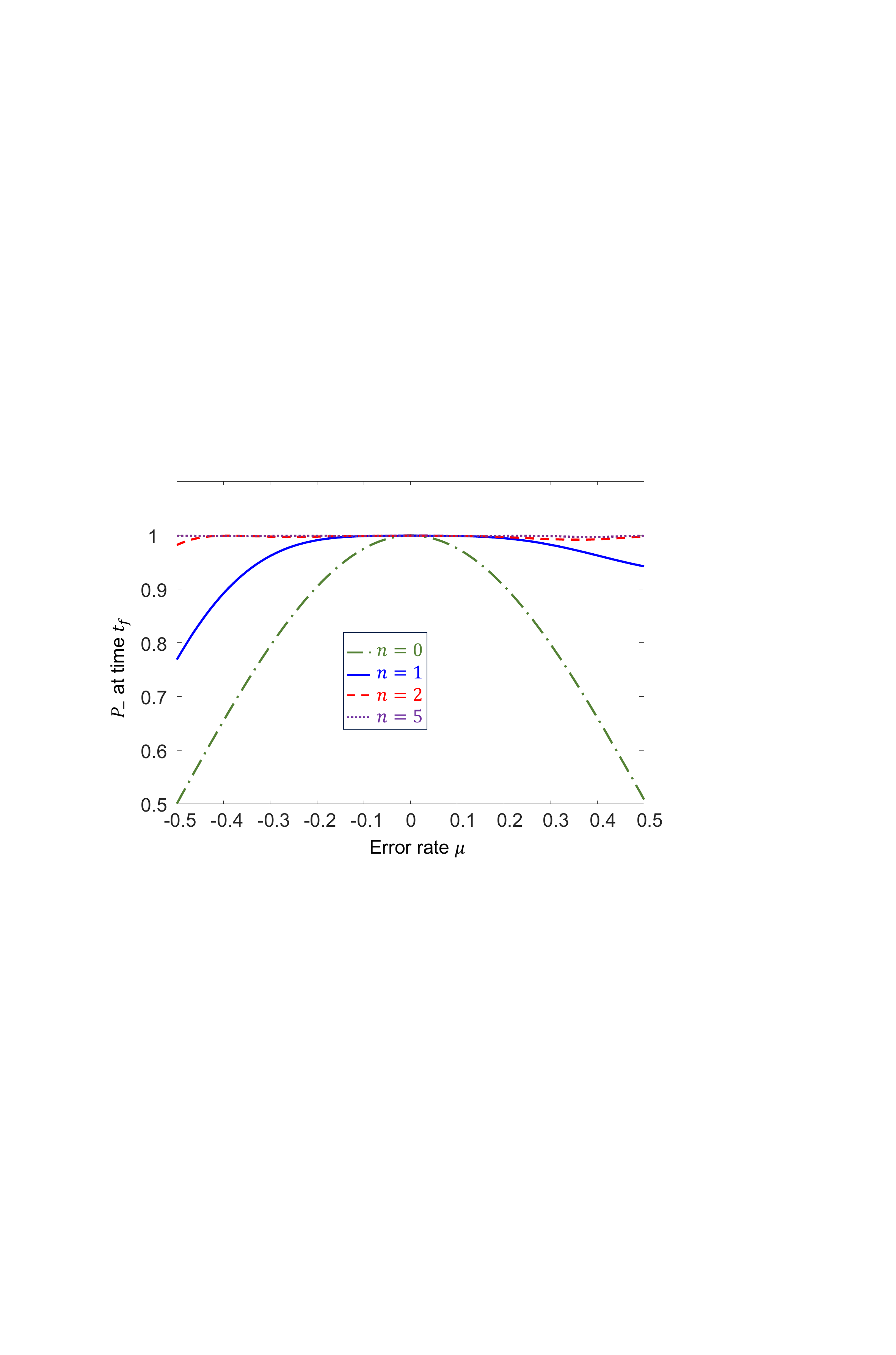}}
	\caption{Population of the odd cat state $|\mathcal{C}_{-}\rangle$ at time $t_{f}$ versus error rate $\mu$ for different control parameters.
		The result of the protocol from Sec.~\ref{sec3} is represented by the green dashed-dotted curve, and the results of the optimal protocol in 
		Sec.~\ref{sec6} is represented by the other curves.
		{The evolution takes place in the complete Hilbert space.}
		}
	\label{fig3}
\end{figure}

    Using perturbation theory up to $O(\mu^2)$, we obtain
    \begin{align}\label{eq26}
        | \psi(t_f) \rangle =& | \psi_0(t_f) \rangle 
                            - {i} \mu \int_{0}^{t_f} dt {U}_0(t_f,t) {H}_1(t) | \psi_0(t) \rangle + \cdots,  
    \end{align}
    where $| \psi_0(t_f) \rangle$ is the solution without perturbation, and ${U}_0(t_{f},t)$ is the unperturbed time evolution operator. We assume that the protocol
    without errors ($\mu = 0$) works perfectly, i.e. $| \psi_0(t_f) \rangle = |\phi_{+}(t_{f}\rangle)=e^{i \beta(t_{f})} | \mathcal{C}_- \rangle$ with real $\beta(t_{f})$. Thus
    \begin{eqnarray*}\label{eq27}
        P_- \approx 1 - \mu^2 \left| \int_{0}^{t_f} dt \langle \phi_-(t) |e^{-iR_{-}} {H}_1(t) e^{iR_{+}}| \phi_+ (t) \rangle \right|^2.
    \end{eqnarray*}
    Substituting the above expression into the Eq.~(\ref{eq25}), we can obtain the systematic error sensitivity
\begin{align}\label{eq28}
        q_s =& \left| \int_{0}^{t_f} dt \langle \phi_-(t) |e^{-iR_{-}} {H}_1(t) e^{iR_{+}}| \phi_+(t) \rangle \right|^2\cr
        =&\frac{1}{4} \left| \int_{0}^{t_f} e^{2i R_+} \Omega_R (-\cos^2 \frac{\gamma}{2} e^{2i \beta} + \sin^2 \frac{\gamma}{2} ) dt \right|^2, 
\end{align}
Using the parameters defined in Sec.~\ref{sec3}, we can numerically calculate Eq.~(\ref{eq28}) and obtain $q_{s}\approx \pi^2/4$.
Note that $\beta$ changes slowly in time [see {Fig.~\ref{fig2}(a)}], we may assume $\beta=-\pi/2$ in Eq.~(\ref{eq28}) and $q_{s}$ can be approximated as
\begin{align}\label{A3}
q_s \simeq \frac{1}{4} \left| \int_{0}^{t_f} \dot{\gamma} dt \right|^2 = \frac{\pi^2}{4},
\end{align}
which coincides {with} numerical result.
The relationship between the population $P_-$ and the systematic error parameter $\mu$ is shown by the
green dashed-dotted curve in Fig.~\ref{fig3}. A deviation rate of $\mu=\pm0.1 $ can lead to an infidelity 
    about $2.5\%$, which is small but 
    causes significant influence in quantum error correction.

    \section{Optimal protocol}\label{sec6}
    Generally speaking, 
    the two-level Hamiltonian for an optimal control protocol \cite{Ruschhaupt2012Njp} takes the following form:
    \begin{align}\label{eq29}
        {H}_{\rm opt} = \frac{1}{2}
        \begin{pmatrix}
            \Delta & {\rm Re}[\Omega] - i {\rm Im}[\Omega]\\
            {\rm Re}[\Omega] + i {\rm Im}[\Omega] & -\Delta
        \end{pmatrix},
    \end{align}
    where ${\rm Re}[*]$ and ${\rm Im}[*]$ denote the real and imaginary parts of the parameter $*$, respectively.
    The derivative of the Lewis-Riesenfeld phases can be obtained through computation
    \begin{align}\label{eq30}
        \dot{R}_{\pm} =\pm \frac{1}{2 \sin\gamma} ( \cos\beta {\rm Re}[\Omega] - \sin\beta {\rm Im}[\Omega]).
    \end{align}
    Using the derivations in Sec.~\ref{sec3}, we can obtain the expressions for ${\rm Re}[\Omega]$, ${\rm Im}[\Omega]$, and $\Delta$ as
	{
	\begin{align}\label{eq31}
        {\rm Re}[\Omega] =&\ 2 \cos\beta \sin\gamma \dot{R}_{+} + \sin\beta\ \dot{\gamma}, \cr
        {\rm Im}[\Omega] =& -2 \sin\beta \sin\gamma \dot{R}_{+} +  \cos\beta\ \dot{\gamma}, \cr
        \Delta =&\ 2 \cos\gamma \dot{R}_{+} - \dot{\beta}.
    \end{align}
	}

    For the Hamiltonian ${H}_{\mathrm{opt}}$ in Eq.~(\ref{eq29}), we can derive {the corresponding} expression for the systematic error sensitivity
    {
	\begin{align}\label{eq32}
        q_s =& \left| \int_{0}^{t_f} dt \langle \psi_-(t) | {H}_1(t) | \psi_+(t) \rangle \right|^2 \cr
		=& \frac{1}{4} \left| \int_{0}^{t_f} dt \left[ 2i e^{2i R_{+}} \dot{R_{+}} \sin\gamma \cos\gamma + e^{2i R_{+}} \dot{\gamma} \right] \right|^2 \cr
		=& \frac{1}{4} \left| \int_{0}^{t_f} dt \left[ e^{2i R_+} \frac{d}{dt} ( \cos\gamma \sin\gamma) + e^{2i R_+} \dot{\gamma} \right] \right|^2. 
    \end{align}
	}
    Note that the boundary values $\gamma(0) = \pi$ and $\gamma(t_f) = 0$, the expression can be further simplified to
    \begin{align}\label{eq33}
        q_s = \left| \int_{0}^{t_f} e^{2i R_+} \dot{\gamma} \sin^2 \gamma\ dt \right|^2. 
    \end{align}
    In the special case where $R_+$ does not vary with time (as in Sec.~\ref{sec5} where $ R_+$ is a constant), we obtain $q_s = {\pi^2}/{4}$.

\begin{figure}
	\centering
	\scalebox{0.42}{\includegraphics{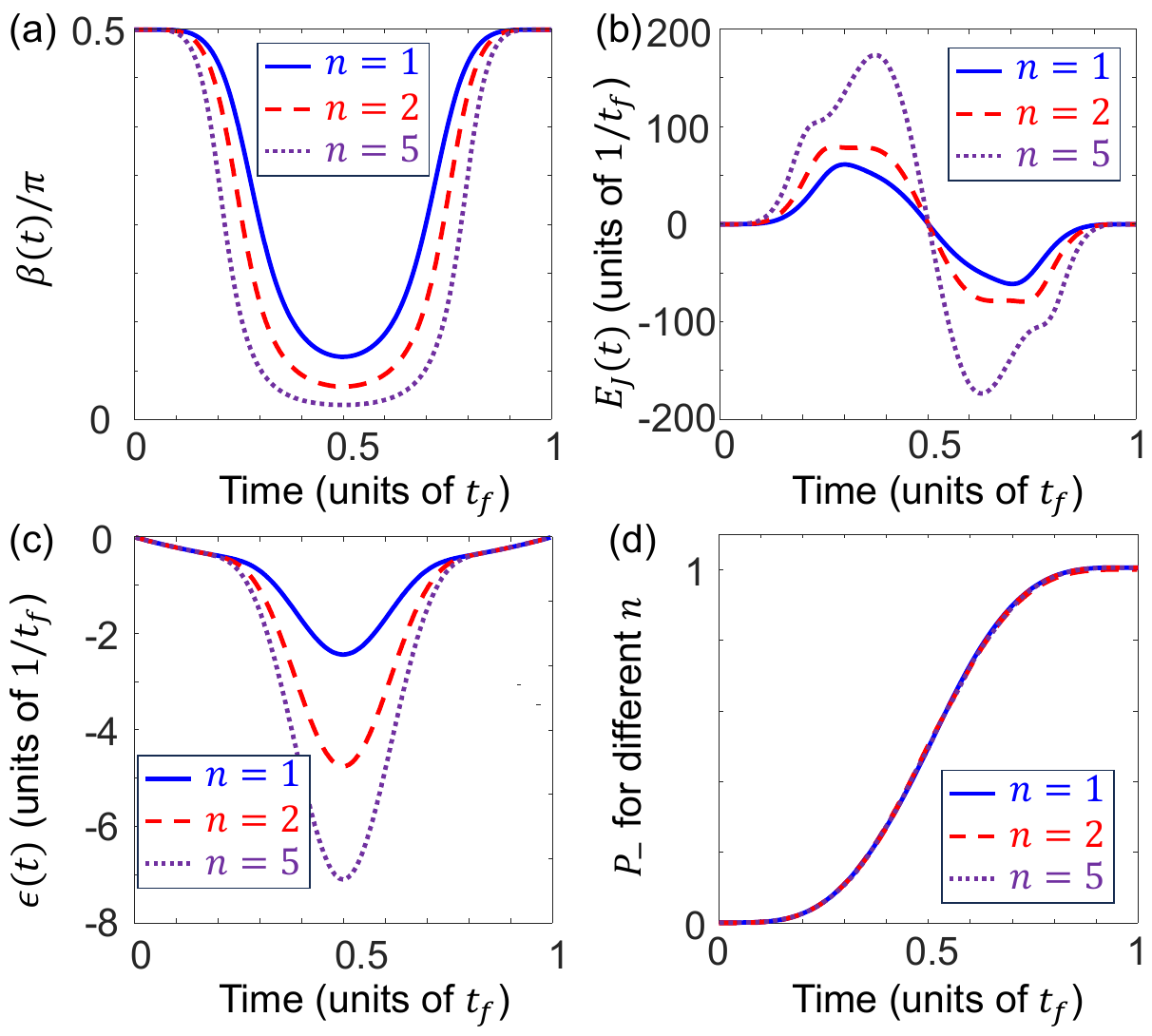}}
	\caption{(a) Parameter $\beta$ calculated according to Eq.~(\ref{eq37}) with the polynomial $\gamma(t) = \sum_{i=0}^{3} a_i t^i$. 
		(b) Corresponding function of $E_J$ calculated by $\Delta$ in Eq.~(\ref{eq36}).
		(c) Corresponding function of $\epsilon$ calculated by $\Omega_{R}$ in Eq.~(\ref{eq36}).
		(d) Time evolution of the odd cat state $|\mathcal{C}_{-}\rangle$ for different $n$. The total evolution time is assumed to be $t_{f}=5/K$.
		{The evolution takes place in the complete Hilbert space.}
		}
	\label{fig4}
\end{figure}

    To make the systematic error sensitivity $q_s = 0$, we consider the case where $R_+$ varies with time\cite{Ruschhaupt2012Njp}, e.g.,
    \begin{align}\label{eq34}
        R_+(t) = \frac{n}{2} (2\gamma - \sin2\gamma),\ \ \ \ \ \ \ (n = 1, 2, 3, \ldots),
    \end{align}
    for $R_+(t)$ in the above expression, we have 
    \begin{align}\label{eq35}
    	q_s = \frac{\sin^2(n \pi)}{4n^2},
    \end{align} 
     so, we have $q_s = 0$ when $n\neq 0$.  Note that in the limit of $n \rightarrow 0$, we 
    obtain $q_s \rightarrow {\pi^2}/{4}$, which is consistent with the previous statement below Eq.~(\ref{eq28}). In this case, the expressions for $\Omega$ and $\Delta$ are as follows
    \begin{align}\label{eq36}
        {\rm Re}[\Omega] =& (4n \cos\beta \sin^3\gamma + \sin\beta) \dot{\gamma}, \cr
        {\rm Im}[\Omega] =& (-4n \sin\beta \sin^3\gamma + \cos\beta) \dot{\gamma}, \cr
        \Delta =& 4n \dot{\gamma} \cos\gamma \sin^2\gamma - \dot{\beta}.
    \end{align}
    For $\Omega$ in Eq.~(\ref{eq31}) to be equivalent with $\Omega_{R}$ in Eq.~(\ref{eq10}), we need to set ${\rm Im}[\Omega]=0$, resulting in
    \begin{align}\label{eq37}
        \cot{\beta}=4n\sin^{3}\gamma.
    \end{align}
    Then, taking this condition in Eq.~(\ref{eq37}) and the boundary condition in Eq.~(\ref{eq19}) and into account,
    we can redesign the parameters $\gamma$ and $\beta$. 
    For instance, we can still use the polynomial expression in Eq.~(\ref{eq20}) for $\gamma$, then we can obtain new $\beta$, as shown at Fig.~\ref{fig4}(a).
    Accordingly, $\Omega$ and $\Delta$ can be calculated by Eq.~(\ref{eq36}), so we can obtain $E_J$ and $\epsilon$, which are shown in Fig.~\ref{fig4}(b) and Fig.~\ref{fig4}(c), respectively.

    Using these optimized parameters, our protocol becomes insensitive to systematic error 
    in the single photon drive (See the blue-solid, red-dashed, and purple-dotted curves in Fig.~\ref{fig3}). 
    As can be seen in Fig.~\ref{fig3}, the systematic error sensitivity can be
    significantly reduced by increase $n$. For $n=5$, a deviation with $\mu=\pm 0.3$ in the parameter $\epsilon$ only
    leads to a decrease of $0.01\%$ in the final population $P_{-}(t_{f})$, resulting in an optimally robust {bit-flip}.
    Noting that the maximums of $|\epsilon|$ and $|E_{J}|$ increase when $n$ increases [see Figs.~\ref{fig4}(b) and (c)],
    a longer operator time $t_{f}$ is needed to satisfy $|\epsilon|,|E_{J}|\ll \omega_{\rm gap}$ for large $n$.
	{ However, optimal control theory (OTC) such as Pontryagin’s Maximum Principle (PMP) \cite{Dridi2020PRL} may be incorporated to minimize 
	the pulse area while maintaining robustness, offering a promising enhancement to our protocol.}
    Figure \ref{fig4}(d) shows the time-dependent population $P_{-}(t)$. It is found that
    increasing $n$ does not change the instantaneous population $P_{-}(t)$. This because the instantaneous population $P_{-}(t)$ is determined by the parameter $\gamma$, which keeps the same for different $n$.

\begin{figure*}
	\centering
	\scalebox{0.47}{\includegraphics{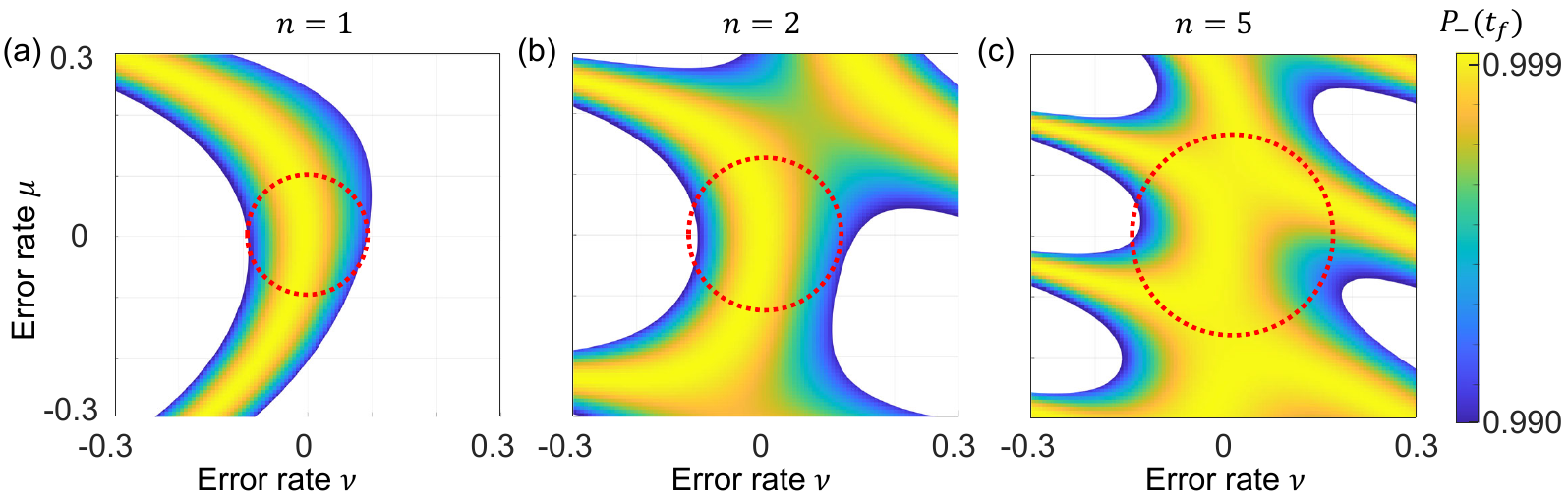}}
	\caption{Population of the odd cat state $|\mathcal{C}_{-}\rangle$ at time $t_{f}$ versus parameter imperfections in $\epsilon$ (with error rate $\mu$) 
	and $E_{J}$ (with error rate $\nu$). 
	{ Within the range of parameter imperfections indicated by the red circle, the fidelity $P_-(t_f)$ remains almost above $99\%$.} 
	Parameters are the same as those in Fig.~\ref{fig4}. The evolution takes place in the complete Hilbert space. 	
	}
	\label{fig6}
\end{figure*}

The above discussion focuses on improving the robustness against parameter imperfections in $\epsilon$. 
When parameter imperfections appear in $E_{J}$, our protocol can also achieve a robust {bit flipping} as shown in Fig.~\ref{fig6}. 
In this case, we consider an additional disturbed Hamiltonian \begin{align}
	H_{2}=-\nu {E_{J}(t)}\{D[i\varphi_{a}\exp(i \omega_{c}t)]+{\rm H.c.}\}/2,
\end{align}
where $\nu$ denotes the error rate.
The population of the target state $|\mathcal{C}_{-}\rangle$ can still reach $\geq 99\%$ when the error rate is $\nu=\pm 0.1$
via our protocol with $n=1$ [see Fig.~\ref{fig6}(a)]. Increasing the value of $n$ can further
improve the robustness against systematic errors. However, to achieve such an optimal
robustness, an increase in the total evolution time is needed as discussed above. This becomes
a defect of the protocol when considering decoherence.
Therefore, for simplicity, we use the pulse with $n=1$ in the following numerical simulations.

    \section{Decoherence}\label{sec8}
\begin{figure}[b]
	\centering
	\scalebox{0.37}{\includegraphics{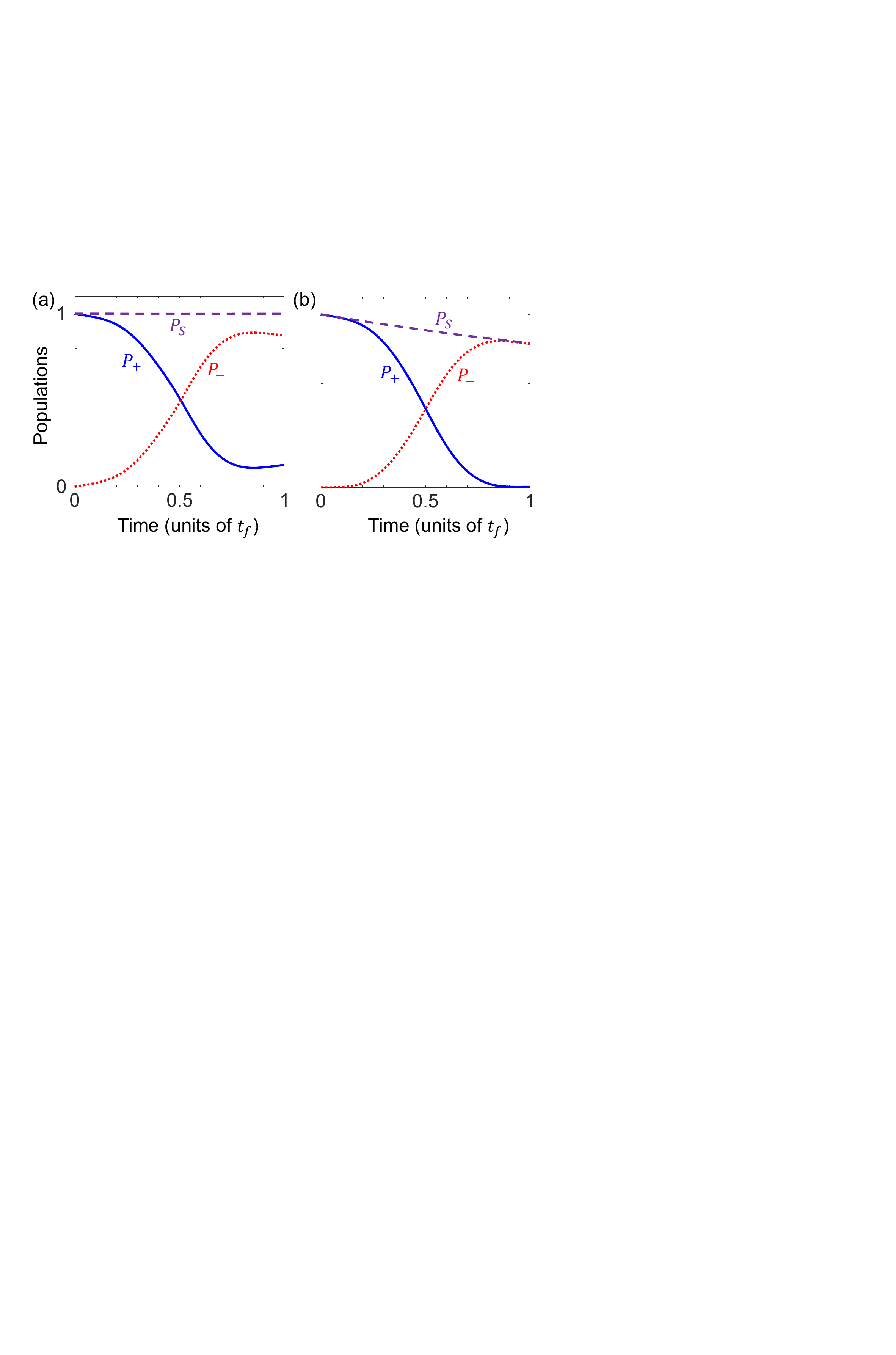}}
	\caption{{Bit flipping} in the presence of (a) single-photon loss and (b) pure dephasing. We choose the optimized protocol with $n=1$ and the total 
	evolution time $t_{f}=5/K$. The single-photon loss and pure dephasing rates are $\kappa=\kappa^{\phi}=0.01K$.
	{The evolution takes place in the complete Hilbert space.}
	}
	\label{fig7}
\end{figure}

    For the resonator, we consider two types of noise: single-photon loss and pure dephasing. The system dynamics are described by the Lindblad master equation \cite{Scully1997Book,Agarwal2012Book}
    \begin{align}\label{eq38}
        \dot{\rho} = -i [{H_{\rm tot}},\rho] + \kappa \mathcal{D}[a] \rho + \kappa^{\phi}\mathcal{D}[a^{\dagger}a] \rho, 
    \end{align}
    where 
    \begin{align}
        \mathcal{D}[o] \rho = o \rho o^{\dagger} - \frac{1}{2} (o^{\dagger}o \rho + \rho o^{\dagger}o) 
        \nonumber
    \end{align}
    is the standard Lindblad superoperator, $\kappa$ is the single-photon dissipation rate, and $\kappa^{\phi}$ is the pure dephasing rate. 
    Projecting the whole system onto the cat-state subspace, we can obtain
    \begin{align}\label{eq40}
    	\dot{\rho}_{\rm eff}\approx & -i[H_{\rm eff},\rho_{\rm eff}]\cr
    	&+\kappa|\alpha|^2\mathcal{D}\left[\frac{A+A^{-1}}{2}\sigma_{x}+ {i} \frac{A-A^{-1}}{2}\sigma_{y}\right]\rho_{\rm eff}\cr
    	&+\kappa^{\phi}|\alpha|^{4}\mathcal{D}\left[\frac{A^{2}+A^{-2}}{2}\mathbbm{1}-\frac{A^{2}-A^{-2}}{2}\sigma_{z}\right]\rho_{\rm eff}, 
    \end{align}
where {$A = \sqrt{\tanh|\alpha|^2}$} 
and $\mathbbm{1}=|\mathcal{C}_{+}\rangle\langle \mathcal{C}_{+}|+|\mathcal{C}_{-}\rangle\langle \mathcal{C}_{-}|$ is the unit matrix in the cat-state subspace. For large $\alpha$, 
$\sigma_{y}$ and $\sigma_{z}$ terms are exponentially suppressed, resulting in
\begin{align}\label{eq41}
	\dot{\rho}_{\rm eff}\approx & -i[H_{\rm eff},\rho_{\rm eff}]+\kappa|\alpha|^2\mathcal{D}[\sigma_{x}]\rho_{\rm eff},
\end{align}
i.e., leaving only the bit-flipping error. This is demonstrated in Fig.~\ref{fig7}(a),
which shows that single-photon loss causes bit-flipping error but no leakage. The sum of populations
\begin{align}
	P_{S}=P_{+}+P_{-},
\end{align}
in the cat-state subspace remains unchanged in the presence of single-photon loss.

According to Eq~(\ref{eq41}), pure dephasing has no influence on the dynamics in the cat-state subspace.
However, pure dephasing can cause leakage out of the cat-state subspace [see Fig.~\ref{fig7}(b)] because 
\begin{align}
a^{\dag}a|\pm\alpha\rangle=|\alpha|^{2}|\pm\alpha\rangle\pm\alpha D(\pm\alpha)|1\rangle.
\end{align} 
This leakage probability is proportional to $|\kappa\alpha/\omega_{\rm gap}|^{2}$ \cite{Puri2019Prx,Puri2020Sa}.
The total population in the cat-state subspace reduces obviously as shown in the figure.

Therefore, considering the full Hilbert space of the cavity mode, the term describing pure dephasing in the master equation for large $\alpha$ should be corrected as
{
\begin{align}\label{R24}
	\mathcal{D}[\sqrt{\kappa^{\phi}}a^{\dag}a]\rho\Rightarrow &\kappa^{\phi}\mathcal{D}\left[P_{\rm{Kerr}}a^{\dag}aP_{\rm{Kerr}}\right]\rho \cr
	\approx &\kappa^{\phi}\alpha^4\mathcal{D}\left[|\mathcal{C}_{+}\rangle\langle\mathcal{C}_{+}|+|\mathcal{C}_{-}\rangle\langle\mathcal{C}_{-}|\right]\rho \cr 
	&{+\kappa^{\phi}\alpha^{2}\mathcal{D}\left[|\psi_{+}^{e,1}\rangle\langle\mathcal{C}_{-}|+|\psi_{-}^{e,1}\rangle\langle\mathcal{C}_{+}|\right]\rho} \cr
	&{+\kappa^{\phi}\alpha^4\mathcal{D}\left[|\psi_{+}^{e,1}\rangle\langle\psi_{+}^{e,1}|+|\psi_{-}^{e,1}\rangle\langle\psi_{-}^{e,1}|\right]\rho}.
\end{align}
}
Here, $P_{\rm Kerr}$ is the projection operator defined as
\begin{align}
	P_{\rm{Kerr}}=|\mathcal{C}_{\pm}\rangle\langle\mathcal{C}_{\pm}|+\sum_{n=1}^{\infty}|\psi_{\pm}^{e,n}\rangle\langle \psi_{\pm}^{e,n}|.
\end{align}
We have ignored the highly excited eigenstates of the { Kerr parametric oscillators (KPOs)} because they are mostly unexcited in the evolution. 
The first line in the right-hand side of Eq.~(\ref{R24}) is a unit matrix in the cat-state subspace.
However, according to the terms in the second line of Eq.~(\ref{R24}), pure dephasing 
can cause transitions from the cat states to the first-excited states, i.e., $|\mathcal{C}_{\pm}\rangle\rightarrow|\psi_{\mp}^{e,1}\rangle$, with a rate $\kappa^{\phi}\alpha^2$.
This causes leakage outside the coding subspace as shown in Fig.~\ref{fig7}.

\begin{figure}[b]
	\centering
	\scalebox{0.36}{\includegraphics{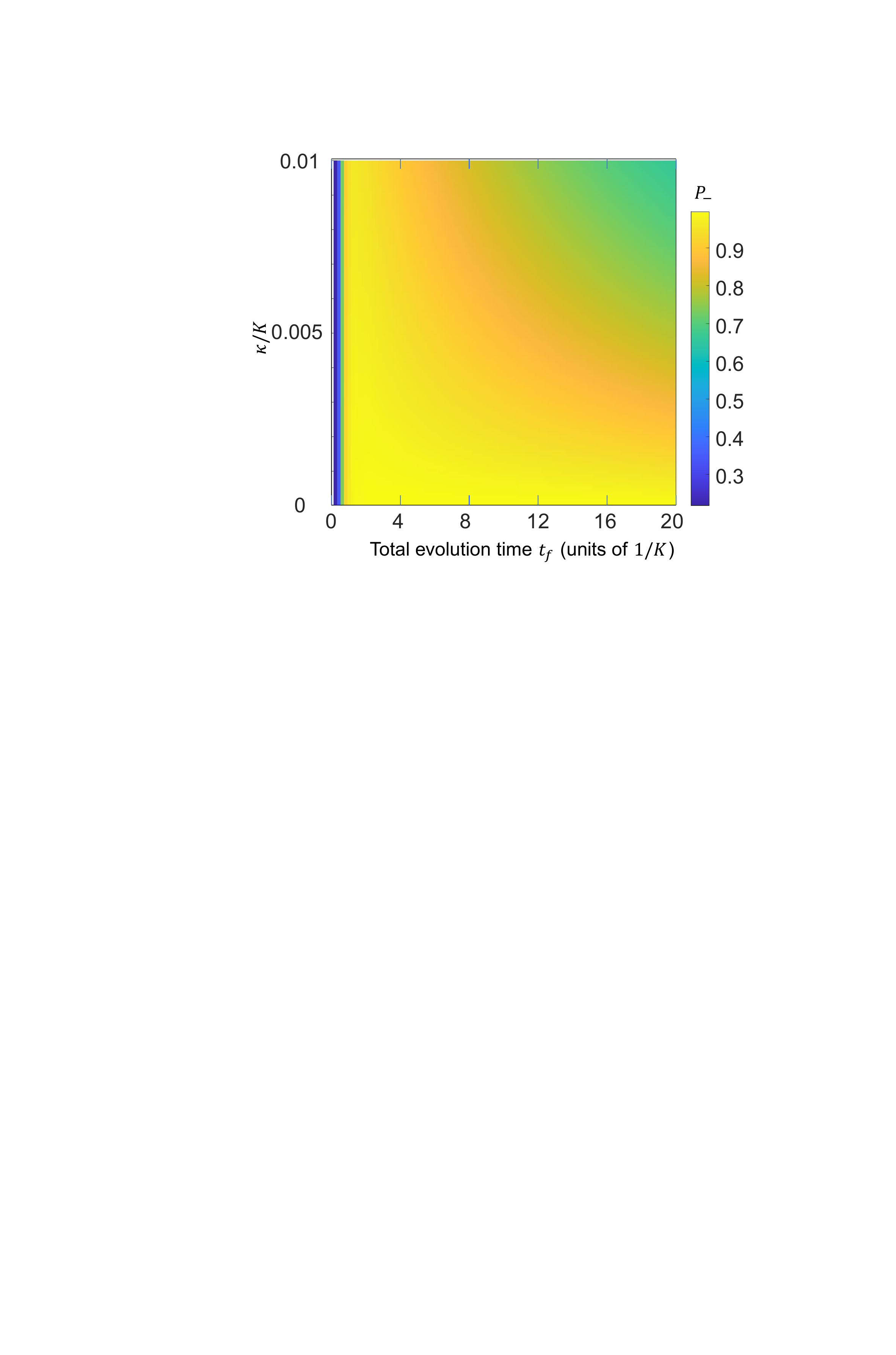}}
	\caption{Population $P_-$ versus the total evolution time $t_{f}$ and single-photon loss rate $\kappa$. We choose the optimized protocol with $n=1$ and 
	assume pure dephasing $\kappa^{\phi}=0$.
	{The evolution takes place in the complete Hilbert space.}
	}
	\label{fig8}
\end{figure}

We also investigate the influence of total evolution time $t_{f}$ on the protocol. As shown in Fig.~\ref{fig8},
an evolution time of $t_{f}\simeq 1/K$ is enough for our protocol to achieve 
the high-fidelity {bit flipping} in the presence of decoherence. For instance, 
when $t_{f}=1.1/K$ and $\kappa=0.01K$, the final population can reach $P_{-}\simeq 96\%$,
demonstrating the effectiveness of our optimized {bit-flipping} protocol.

\section{Discussion and Conclusions}\label{sec9}

The proposed protocol is possible to be realized in superconducting
circuits {as discussed in Appendix \ref{secA1}} \cite{Koch2007Pra,You2007,Flurin2015,Grimm2020,Wustmann2013,Gu2017,Krantz2019APR,Kjaergaard2020,Kjaergaard2020Arc,Kwon2021Jap}, especially the platforms with 3-dimension cavities \cite{Cai2021Fr,Ma2021Sb,Hu2019Np,Xu2020Prl,Cai2021Prl,Wang2022Nc,Ni2023Nat}.
This is because that the 3-dimension cavities can provide
a relatively long coherent time (extending to microseconds) for photonic qubits \cite{Cai2021Fr,Ma2021Sb}.
Cat-state qubits also belong to a larger family of bosonic qubits, most of which have
been realized with 3-dimension cavities \cite{Cai2021Fr,Ma2021Sb}.
To be specific,
the Kerr nonlinearity and the two-photon drive can be respectively realized
by the Josephson junction (transmon) nonlinearity and four-wave mixing \cite{Chen2019Pra,Qin2019Pra,Qin2020NanoP,Qin2021,Puri2019Prx}.
The control Hamiltonian in Eq.~(\ref{eq5}) is possible to realize
by {capacitively coupling} the Kerr-nonlinear 
mode to a
Josephson junction and assuming that other modes (including the junction mode) are never excited \cite{Cohen2017Prl}. 
Following the first cat-state qubit experiment, we can consider the experimental 
parameters $K/2\pi=6.7~$MHz, $\kappa/2\pi\simeq 0.01~$MHz, and $\kappa^{\phi}/2\pi\simeq0.045~$MHz.
With these parameters and $t_{f}=1.1/K\simeq 26$~ns, the final population of the target state is $P_{-}(t_{f})\simeq 95\%$ 
in the presence of parameter imperfection with $\mu=\nu=0.1$ and decoherence.
{
Table \ref{Tab1} presents the comparison between the protocol we proposed and existing alternatives, our proposed protocol offers several unique 
strengths that distinguish it from these alternatives: By extending the two-level system in Conventional-STA approaches to the Kerr-nonlinear system, we achieve 
high-fidelity {bit flipping} without relying on complex numerical optimization. Our method explicitly incorporates the Kerr-nonlinear spectrum, ensuring 
enhanced robustness against both leakage and parameter drifts. And the smooth control pulses are experimentally feasible with current hardware capabilities.
A possible generalization of our protocol to the two-qubit control not gate is given in Appendix \ref{secA2}. 
\begin{table}
	\centering
	\begin{tabular}{|p{4cm}|p{4cm}|p{4cm}|}
	\hline
	{\textbf{Method}} & {\textbf{Advantages}} & {\textbf{Limitations}} \\ \hline
	{Robust STA (This Work)} & {Fast, robust control; leakage suppression} & {Requires careful parameter selection} \\ \hline
	{Optimal Control Theory \cite{Machnes2011PRA}} & {Achieves optimal fidelity; flexible design} & {Complex pulse shapes; computationally demanding} \\ \hline
	{Dynamical-Decoupling \cite{Piltz2013PRL}} & {Protection against certain decoherence types} & {Limited protection against bit-flip errors} \\ \hline
	\end{tabular}
	\caption{{Comparison with existing control techniques}}
	\label{Tab1}
\end{table}

In conclusion, we have investigated a feasible control method 
to obtain the optimally robust shortcut to {state transfer} in cat-state qubits.
Focusing on the Kerr-cat qubit, which is realized by parametrically driving a Kerr-nonlinear resonator,
we have constructed shortcuts to adiabatic passages and minimized the systemic error
sensitivity based on the invariant-based reverse engineering.
{It is worth noting that another equally popular method for stabilizing cat qubits is to use two-photon 
dissipation \cite{Gautier2022PRXQ, Leghtas2013PRA, Leghtas2015Science, Touzard2018PRX, Reglade2024Nature, Gautier2023PRXQ}, 
and this protocol we proposed may be possible to be extended to such two-photon dissipation schemes in future work.}
Future work will involve extending our results to other logic qubits and the multi-qubit cases.
The existence of a set of optimal solutions for systematic errors also opens the
way to further optimization with respect to other error-correcting qubits.

\section*{Acknowledgements}
	Y.-H.C. was supported by the National Natural Science Foundation of China under Grant No. 12304390 and 12574386, the Fujian 100 Talents Program, and the Fujian Minjiang Scholar Program. Y. X. was supported by the National Natural Science Foundation of China under Grant No. 62471143, the Key Program of National Natural Science Foundation of Fujian Province under Grant No. 2024J02008, and the project from Fuzhou University under Grant No. JG2020001-2.

{
\appendix
	\section{Hamiltonian of a parametric oscillator}\label{secA1}
	The Hamiltonian of the SQUID array resonator shown in Fig.~\ref{fig0}(a) is 
	\begin{align}
		H_{K}^{0}=4E_{C}\hat{n}^2-NE_{J}(\Phi(t))\cos\left(\frac{\hat{\phi}}{N}\right),
	\end{align}
	where $\hat{n}$ is the number of Cooper pairs and $\hat{\phi}$ is the overall phase across the junction array. $E_{C}$ and $E_{J}$
	are the resonator's charging energy and the Josephson energy for a single SQUID, respectively. $N$ is the number of SQUIDs in the array.
	$\Phi(t)$ is an additional flux for controlling the Josephson energy $E_{J}$.
	
	We assume that the Josephson energy $E_{J}$ is modified as (with a frequency $\omega_{p}$)
	\begin{align}
		E_{J}[\Phi(t)]=E_{J}+\delta E_{J}\cos(\omega_{p} t).
	\end{align}
	After applying the Taylor expansion of $\cos\left(\hat{\phi}/N\right)$ to
	fourth order, we obtain
	\begin{align}
		H_{K}^{0}\approx &~ 4E_{C}\hat{n}^2-NE_{J}(1-\hat{X}+\hat{X}^2/6)\cr
		&- N\delta E_{J}(1-\hat{X})\cos(\omega_{p}t) 
	\end{align}
	where $\hat{X}=(\hat{\phi}/N)^2/2$. The highest level is assumed to be much smaller than the dimension of the Hilbert space.
	Following the standard quantization
	procedure for circuits \cite{Koch2007Pra,You2007}, we can define ($\hbar=1$)
	\begin{align}
		\hat{n}=-i n_{0}(a-a^{\dag}), \ \ \ \ \ \ \ \ \ \ \ \ \ \ \ \hat{\phi}=\phi_{0}(a+a^{\dag}),
	\end{align}
	where $n_{0}=\sqrt[4]{E_{J}/(32NE_{C})}$ and ${\phi_{0}}=2\sqrt{2}/n_0$ are the
	zero-point fluctuations.
	The Hamiltonian $H_{K}^{0}$ becomes 
	\begin{align}\label{R37}
		H_{K}^{0}=&~\omega_{c}a^{\dag}a-\frac{E_{C}}{12N^{2}}\left(a+a^{\dag}\right)^{4}\cr&+\frac{\delta E_{J}\omega_{c}}{4E_{J}}\left(a+a^{\dag}\right)^2 \cos(\omega_{p}t),
	\end{align}
	where $\omega_{c}=\sqrt{8E_{C}E_{J}/N}$. Here, we have dropped the constant terms for simplicity.
	Therefore, by transforming the Hamiltonian into a rotating frame at the frequency $\omega_p/2$, we can neglect 
	all fast-oscillating terms by the rotating-wave approximation, resulting in
	\begin{align}
		H_{\rm Kerr}=-Ka^{\dag 2}a^{2}+P(a^{2}+a^{\dag 2}),
	\end{align}
	where $K=E_{C}/N^2$ and $P=(\omega_{c}+K)\delta E_{J}/8 E_{J}$.
	
{
\section{Coupling a Kerr-nonlinear mode to a Josephson junction}\label{EJ}
As illustrated in Fig.~\ref{fig0}(b), we consider a Kerr-nonlinear mode with frequency $\omega_c$ that is 
capacitively coupled to a Josephson junction. Assuming that other modes (including the junction mode) remain unexcited, the Hamiltonian in the interaction picture reads
\begin{align}
	H_{\rm int}(t) &= -E_J(t) \cos[\varphi_a(a e^{-i \omega_c t} + a^\dagger e^{i \omega_c t} )] \cr
				   &= -\frac{E_J(t)}{2} \{ D[i \varphi_a e^{i \omega_c t}] + D[-i \varphi_a e^{i \omega_c t}] \}. 
\end{align}
Expanding the displacement operator, 
\begin{align}
	D[i \varphi_a e^{i \omega_c t}] = \sum_{l_a = 0}^{\infty} \mathcal{A}(l_a)(-a e^{-i \omega_c t})^{l_a} + \sum_{l_a = 1}^{\infty} (a^\dagger e^{i \omega_c t})^{l_a} \mathcal{A}(l_a), 
\end{align}
where $\mathcal{A}(l_a)=\varphi_a^{l_a} e^{-\varphi_a^2/2} \sum_{n_a=0} \frac{n_a!}{(n_a+l_a)!} L_{n_a}^{(l_a)}(\varphi_a^2)|n_a\rangle\langle n_a|$ is a hermitian 
operator, and $L_{n_a}^{(l_a)}(*)$ is the generalized Laguerre polynomial of order $n_a$ and parameter $l_a$. For $E_J(t) \ll \omega_c$, under the rotating-wave 
approximation, the interaction Hamiltonian becomes 
\begin{align}
	H'_{\rm int}(t)=E_J(t)e^{-\varphi_{a}^{2}/2}\sum_{m=0}^{\infty}L_{m}(\varphi_{a}^{2})|m\rangle\langle m|, 
\end{align}
where $L_{m}(*)$ is the Laguerre polynomial of order $m$.
}

	\section{The flipping of the coherent states}\label{Re}
	According to Eq. (\ref{eq3}), in the limit of large $\alpha$, the coherent states $|\pm \alpha \rangle$ can be given by 
	\begin{align}\label{Re1}
		|\pm \alpha \rangle \simeq \frac{\sqrt{2}}{2} (| \mathcal{C}_+ \rangle \pm | \mathcal{C}_- \rangle ),
	\end{align}
	to achieve the flipping of the coherent states $|\pm \alpha \rangle$, we redesign the boundary conditions of $\gamma$ and $\beta$, 
	\begin{align}\label{Re2}
        &\gamma(0) = \frac{\pi}{2}, \ \ \ \gamma(t_f) = \frac{\pi}{2}, \ \ \ \dot{\gamma}(0) = \dot{\gamma}(t_f)= 0, \cr
        &\beta(0) = 0, \ \ \ \ \beta(t_f) = \pi, \ \ \ \ \dot{\beta}(0) = \dot{\beta}(t_f)= 0.
    \end{align}
	To satisfy the boundary conditions given in Eqs.~(\ref{Re2}), we can assume 
	\begin{align}\label{Re3}
		\gamma(t) = \frac{\pi}{2}, \ \ \ \ \  {\rm and}  \ \ \ \  \ \beta(t) = \sum_{i=0}^{3} b_i t^i,
	\end{align}
	and thus determine their values as shown in Fig. \ref{R1}(a). Accordingly, we can obtain $E_J$ and $\epsilon$. Such parameters allow 
	a {bit flipping} from the coherent state $|\alpha\rangle$ to $|-\alpha\rangle$ through a nonadiabatic passage.

	In Fig.~\ref{R1}(b), we display the dynamical evolution of the system when the initial state is $|\alpha\rangle$.
    An almost perfect {bit flipping} ($P_{-}\simeq 99.9\%$ at $t=t_{f}$) is obtained as shown in the figure, where
    the populations of the states $|\alpha\rangle$ and $|-\alpha\rangle$ are defined as 
    \begin{align}\label{Re4}
      P_{\pm}(t)=|\langle \pm \alpha|\psi(t)\rangle|^{2}.
    \end{align}

	Since $\gamma(t)$ is a constant, and $\Omega_R = \dot{\gamma} / \sin{\beta}$, so $\Omega_R = 2 (\alpha^* + \alpha) \epsilon = 0$, that is, $\epsilon=0$. 
	Consequently, the disturbed Hamiltonian is given by $H_{1}=\epsilon(\alpha^{*}+\alpha)\sigma_{x}=0$, so the systematic error sensitivity $q_{s} = 0$.

	\begin{figure}
		\centering
		\scalebox{0.6}{\includegraphics{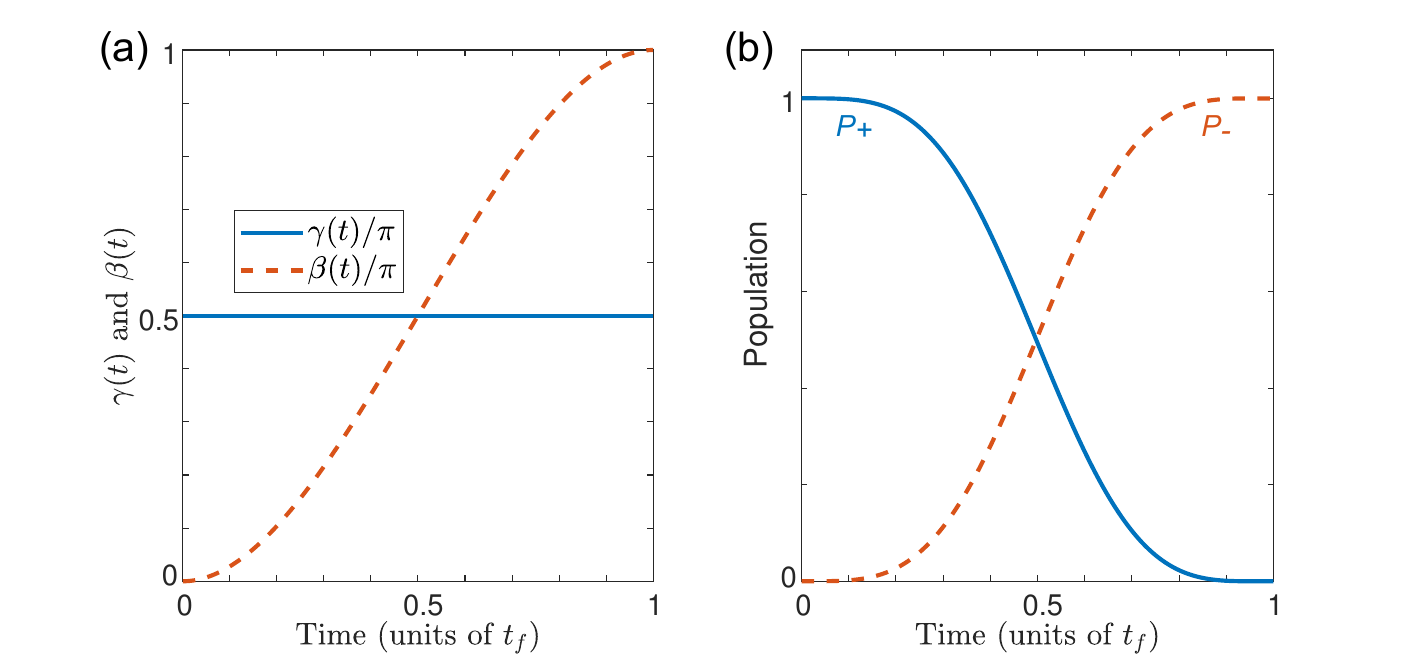}}
		\caption{(a) Parameters for STA method $\gamma(t) = \frac{\pi}{2}$ and $\beta(t) = \sum_{i=0}^{3} b_i t^i$. 
			(b) Non-adiabatic {state transfer} in the coherent states. 
			The evolution takes place in the complete Hilbert space.}
		\label{R1}
	\end{figure}

	\section{Two-qubit CNOT gate}\label{secA2}
The generalization of our protocol to the multiqubit cases is not a difficult work.
For instance, the two-qubit control-not (CNOT) gate is defined as
\begin{align}
	U_{\rm{CNOT}}=\frac{1}{2}\mathbbm{1}\otimes\left(\mathbbm{1}+\sigma_{z}\right)+\frac{1}{2}\sigma_{x}\otimes\left(\mathbbm{1}-\sigma_{z}\right),
\end{align}
where $\mathbbm{1}$ is the unit operator of a two-level system.
Such a gate can be implemented with the effective evolution operator 
\begin{align}
	U_{\rm eff}(t)=\frac{1}{2}\mathbbm{1}\otimes\left(\mathbbm{1}+\sigma_{z}\right)+\frac{1}{2}U_{x}(t)\otimes\left(\mathbbm{1}-\sigma_{z}\right),
\end{align}
where $U_{x}(t)$ is the evolution operator of the single cat-state qubit in our manuscript, satisfying $U_{x}(0)=\mathbbm{1}$ and $U_{x}(t_{f})=\sigma_{x}$.
Therefore, the parameters used for the single-qubit gate in the manuscript can be directly applied to the two-qubit case.
Hence, when $U_{x}(t_f)=\sigma_x$, 
the evolution operator ${U}_{\rm{eff}}(t_f)$ corresponds to a CNOT gate.

Now, the problem is, how to encode the second qubit? Encoding on cat states or not?
It is well known that cat-state qubits are an important kind of bosonic qubits. 
Such qubits are a promising candidate for realizing fault-tolerant quantum computing because they have a biased noise 
channel that the bit-flip error dominates over all the other errors. 
For our protocol to preserves the
error bias, the dominant error operators should commute with the evolution operator. 
For the first qubit, the dominant error operator is $\sigma_{x}$ [see Eq. (\ref{eq40}) in the revised manuscript].
It commutes with the gate operator $U_{\rm CNOT}$. However, if the second qubit is also a cat-state qubit, 
the gate operator $U_{\rm CNOT}$ cannot commute with the error operator $\sigma_{x}$, but can commute with an error operator $\sigma_{z}$.
That is, for the protocol to preserves the
error bias, the dominant error operator for the second qubit should be the operator $\sigma_{z}$.
This is possible by encoding the second qubit on coherent states $|\pm\beta\rangle$.
That is, the Pauli matrices of the second qubit should be
\begin{align}
	\sigma_{x}=&|\beta\rangle\langle -\beta|+|-\beta\rangle\langle \beta|,\cr
	\sigma_{y}=&-i|\beta\rangle\langle -\beta|+i|-\beta\rangle\langle \beta|,\cr
	\sigma_{z}=&|\beta\rangle\langle \beta|-|-\beta\rangle\langle -\beta|.
\end{align}

Based on the evolution operator ${U}_{\rm{eff}}(t)$, we can reversely deduce the corresponding effective Hamiltonian as 
\begin{align*}
	{H}_{\rm{eff}}^{(2)}(t)=&i\dot{{U}}_{\rm{eff}}(t){U}_{\rm{eff}}^{\dag}(t)\cr
	=&\frac{1}{2}H_{\rm{eff}}(t)\otimes\left(\mathbbm{1}-\sigma_{z}\right)\cr
	=&\frac{1}{2}H_{\rm{eff}}(t)\otimes\left[\mathbbm{1}-\left(|\beta\rangle\langle \beta|-|-\beta\rangle\langle -\beta|\right)\right],
\end{align*}
where $H_{\rm eff}(t)$ is the Hamiltonian in Eq. (\ref{eq10}) in the manuscript. 
This effective Hamiltonian is simplified from
\begin{align}
	H(t)=&H_{\rm Kerr}^{(a)}+H_{\rm Kerr}^{(b)}\cr
	     &+\frac{1}{2}H_{c}(t)\otimes\left[\mathbbm{1}-\beta(b+b^{\dag})\right],
\end{align}
where $b$ ($b^{\dag}$) is the annihilation (creation) operator
of the second cavity mode.

Similar results can
be found applying the same methods to different two-qubit gates.
The most important thing is, one needs to find the right codes so that the protocol can still be fault-tolerant.
}

\section*{Author contributions}
Y.-H.C. conceived and developed the idea. S.-W.X., J.-T. Y., K.-X. Y, and Z.-Z.Z. analyzed the data and
performed the numerical simulations, with help from Y.-H.C. and Y.X.. S.-W.X., Z.-Z.Z., Z.-Y. Z. and Y.-Y.G. cowrote the paper with
feedback from all authors.

\section*{Data availability}

The data used for obtaining the presented numerical results as well as for generating the plots is available on request. Please refer to yehong.chen@fzu.edu.cn

\section*{Competing interests}

The authors declare that they have no competing interests.
    

\end{document}